\documentstyle[aps,psfig,multicol]{revtex}
\newcommand{\cprime}{{\cal{C}^{\prime}}}
\newcommand{\cdprime}{{\cal{C}^{\prime\prime}}}

\newcommand{\ndp}{n^{\prime\prime}}
\newcommand{\ccp}{\cal{C} \rightarrow \cprime}
\newcommand{\cdc}{\cdprime \rightarrow \cal{C}}

\newcommand{\rt}{\tilde\rho}

\newcommand{\htx}{\partial_x h}
\newcommand{\htxx}{ \partial_{xx} h}
\newcommand{\htt}{ \partial_t h}

\begin{document}
\title{Steady State and Dynamics of  Driven Diffusive Systems with
       Quenched Disorder}

\author{Goutam Tripathy and Mustansir Barma \\
Tata Institute of Fundamental Research \\
Homi Bhabha Road, Mumbai 400 005, India}

\maketitle

\begin{abstract}
We study the effect of quenched disorder on nonequilibrium systems of
interacting particles, specifically, driven diffusive lattice gases with
spatially disordered jump rates. The exact steady-state measure is found
for a class of models evolving by drop-push dynamics, allowing several
physical quantities to be calculated. Dynamical correlations are studied
numerically in one dimension.  We conjecture that the relevance of quenched
disorder depends crucially upon the speed of the kinematic waves in the
system. Time-dependent correlation functions, which monitor the dissipation
of kinematic waves, behave as in pure system if the wave speed is
non-zero. When the wave speed vanishes, e.g. for the disordered exclusion
process close to half filling, disorder is strongly relevant and induces
separation of phases with different macroscopic densities. In this case the
exponent characterizing the dynamical correlation function changes.

Pacs numbers: 05.60.+w, 05.40.+j, 05.50.+q, 05.70.Ln
\end{abstract}

\begin{multicols}{2}
What is the effect of quenched disorder on driven, nonequilibrium
systems?  This question is important in a number of physical
situations involving flow in random media \cite{Fisher}.
Theoretically, our understanding of these phenomena is based largely
on numerical simulations and on the analysis of continuum equations
for coarse-grained variables.  In this Letter we obtain the exact
steady state and static correlation function for a class of models of
driven, interacting particles on a lattice -- the drop-push process --
with quenched site disorder in the hopping rates.  Further, we study
the time-dependence of hydrodynamic fluctuations for this system in
one dimension, and also for the disordered asymmetric exclusion
process.  We find that the behaviour of a system with current $J_0$
and density $\rho$ is largely determined by $c_0=\partial J_0/\partial
\rho$. If the density is uniform on a macroscopic scale, $c_0$ is the
mean speed of kinematic waves which transport density fluctuations
through the system \cite{LW}. If $c_0$ is nonzero, we conjecture that
quenched disorder does not affect the asymptotic behaviour of the
decay of fluctuations, and support this with extensive numerical
results. By contrast, we find that vanishing $c_0$ can signal the
onset of disorder-induced phase separation with coexisting macroscopic
regions of different density, in which case the dynamical behaviour is
different.

In a coarse-grained description of a 1-$d$ disordered current-carrying
system, the nonuniform steady state density profile is described by a
function $\rho_0(x)$.  The evolution of density fluctuations
$\rt\equiv\rho(x,t)-\rho_0(x)$ is described phenomenologically by a
stochastic evolution equation with spatially random coefficients:
\begin{equation}
\partial_t\rt=
\partial_x [D(x) \partial_{x}\rt- c(x)\rt-\lambda(x)\rt^2...  -\eta(x,t)].
\label{delrho}
\end{equation}
This follows from the continuity equation 
$\partial_t\rt(x,t)+\partial_xJ(x,t)=0$ 
on writing the current as 
$J(x,t)=J_{sys}(x,t)-D(x)\partial_x\rt+\eta(x,t)$, where $D(x)$ is
the space-dependent diffusion constant and $\eta$ is white noise; the
systematic part of
the current $J_{sys}$ is expanded as $J_0+c(x)\rt+\lambda(x)\rt^2$....
The most relevant source of disorder in (\ref {delrho}) is 
the term with coefficient $c(x)$
which represents the space-varying local speed 
of the kinematic wave of density fluctuations. The problem is equivalent to
that of a moving interface in the presence of a certain type of 
columnar disorder; the interface height  $ h(x,t)$ is related to
$\rt$ by $\rt=\partial_x h$. Equation (1) then becomes
\begin{equation}
\htt = D(x)\htxx -c(x) \htx -\lambda(x) (\htx)^2...  - \eta(x,t).
\label{delh}
\end{equation} 
The random coefficients $c(x),\lambda(x)$ now represent columnar
disorder in the 2-$d$ $h$-$x$ space. It is important to
understand how disorder affects scaling properties and
to identify factors responsible for different universality classes
\cite {Amaral}. A crucial difference between
(\ref{delh}) and the model studied by
Krug \cite {Krug} is the absence of an additive quenched noise
$\epsilon(x)$ which models a frozen random contribution to interface
mobility. Such a term, which strongly influences static and
dynamic properties, is necessarily absent in (\ref{delh}) as a
consequence of the spatial constancy of $\langle J_{sys}\rangle$ implied by
particle conservation. Further, our model is distinct from depinning-
threshold interface models with quenched {\it point} disorder \cite{Amaral}.
We show that the roughness exponent $\alpha=1/2$ in contrast
to \cite{Amaral} and \cite{Krug}.
The decay of the kinematic wave in time leads to a
power-law $\sim t^{2\beta}$ growth of the correlation function
\begin{equation}
S^2(t) \equiv \langle[h(x+c_0t,t)-h(x,0)-J_0t]^2\rangle.
\label{S2}
\end{equation}
We conjecture that as long as $c_0$ is nonzero, leading power law behaviours
are the same as in the pure system with no disorder ($\beta={1\over 3}$)
\cite{vanB,KPZ}. 
This is supported by Monte Carlo simulation results for the
disordered drop-push process, and for the disordered exclusion process in
the regime $|\rho-1/2| > \Delta$ with $\Delta \ne 0$.
For $|\rho-1/2| < \Delta$ we present evidence that $J_0$ is
independent of $\rho$.  In this regime, $c_0$ vanishes and shocks separate
macroscopic regions of different mean densities.

The disordered drop-push process (DDPP) is a model of driven transport
of carriers trapped in local regions of space, with each move
involving a cascade of transfers through filled traps \cite{BR,SRB}.
In 1-$d$, on every site $i$ of 
an $L-$site ring is a well which can accommodate at most $l_i$ 
\begin{figure}
\psfig{figure=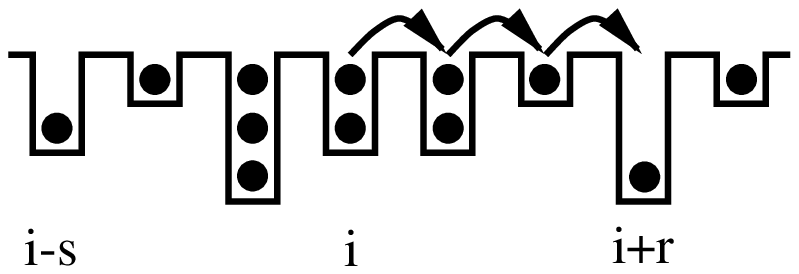,clip=true,width=8cm}
Fig. 1~ Typical DDPP configuration and move.
\label{fig1}
\end{figure}
\noindent particles (Fig. 1). Each well depth $l_i$ is chosen independently from a
probability distribution $P(l_i)$.  The configuration of the system is
specified by the set of particle occupation numbers $\{n_i\}$ of all
the wells. The dynamics is stochastic, and in a small time interval
$dt$, there is a probability $\epsilon(n_i|l_i) dt$ that a particle
from well $i$ hops out, and drops into well $i+1$. If well $i+1$ is
already full, a particle from this well now gets pushed further right,
and so on (Fig. 1).  The cascade of adjacent-site jumps terminates
once a particle drops into a well (say $i+r$) which was not completely
full earlier.  This elementary move thus changes configuration $\cal C
\equiv \{....$ $ n_i~ l_{i+1}~ l_{i+2}~ ...l_{i+r-1} ~n_{i+r}~ ....\}$
to $\cal C'
\equiv \{....$ $n_i-1~ l_{i+1}~
l_{i+2}~...l_{i+r-1}~n_{i+r}+1~....\}$.  The jump rates
$\epsilon(n_i|l_i)$  are pre-specified, and depend both on the depth
of the well and the occupation. The rates and well depths are 
quenched random variables.

The probability  for the system
to be in configuration $\cal{C}$ satisfies the standard master equation
\cite{MEQ} with the transition probabilities $W({\ccp})$ identified
with $\epsilon$'s above.  
In the steady state, the total flux out of $\cal{C}$ equals the total
incoming flux. This is ensured if
for every $\cprime$ obtained from ${\cal C}$ by an elementary
transition there is a unique configuration $\cdprime$ such that in the
steady state
\begin{equation}
W({\ccp}) \mu({\cal C}) = W({\cdc}) \mu(\cdprime).
\label{pwb}
\end{equation}
Here $\mu(\cal C)$ is the invariant measure and (\ref{pwb}) is the  
condition of pairwise balance \cite{SRB}.
We define weights for single-site occupations by $u_i(0)=1$ and
$ u_i(n_i)= \tau_i(1)~\tau_i(2)\cdots\tau_i(n_i) ~~{\rm if}~
0<n_i\le l_i.
$
with $\tau(n_i)\equiv \epsilon_0/\epsilon(n_i|l_i)$, where
$\epsilon_0$ is a microscopic rate.
The measure for 
configuration ${\cal C} \equiv \{ n_i\}$ has the product form
\begin{equation} \mu({\cal C})=\prod_{i=1}^L u_i(n_i). 
\label{InvM}
\end{equation}
To show that this satisfies (\ref {pwb}), we construct the
configuration $\cdprime$ corresponding to a given $\cal C$ and
$\cprime$ as follows.  Suppose the transition ${\cal C}\rightarrow
\cprime$ involves hopping a
particle at well $i$ to well $i+r$ with all wells in between full. Also
suppose well $i-s$ is not full but all wells between $i-s$ and $i$ are full
(Fig. 1). Configuration $\cdprime$ is identical to $\cal C$ at all sites
except at the sites $i-s$ and $i$, at which $\ndp_{i-s}=n_{i-s}+1,~ \ndp_i =
n_i-1$. Then (\ref{pwb}) is satisfied, in view of the measure
(\ref{InvM}). Since the dynamics is ergodic the invariant measure
(\ref{InvM}) is unique for a fixed number of particles \cite{MEQ}. This
generalizes the result obtained earlier for the non-disordered case
\cite{SRB}, and is the first instance of an exact determination of the
steady state of a spatially disordered nonequilibrium system of
interacting particles \cite{traffic}.

  The product measure form (\ref{InvM}) allows us to calculate several
physical quantities: the site
densities, height-height correlation function, current $J_0$, and
wave speed $c_0$. In the limit of large $L$, it is convenient to introduce the
generating function $Z_i=\sum^{l_i}_{n_i=0} u_i(n_i)z^{n_i}$ where $z$ is
the fugacity.  The probability of occurrence of configuration ${\cal C}$ is
${\cal P}({\cal C})= \mu({\cal C}) z^{N_P}/{\prod_i Z_i}$, where $N_P$ is
the number of particles in the configuration.  The steady state is
characterized by uniform $z$, but inhomogeneous site densities $\langle
n_i\rangle = z \partial ln Z_i/\partial z$.
The height-height correlation
function $\Gamma^2(r)\equiv\langle (h_{i+r}-h_i)^2\rangle $ becomes
$\sum_{j=i+1}^{i+r} (\langle n_j^2\rangle-\langle n_j\rangle^2)$ and  can be
evaluated. Disorder averaging gives $\Gamma(r) \sim r^{1/2}$, implying
$\alpha=1/2$.

To find the steady state current $J_0$, note that the current
$J_{i,i+1}$ across bond $(i,i+1)$ comes from jumps which either $(a)$
originate from sites to the left of site $i$ (with in-between wells
full), or $(b)$ originate from site $i$ itself (a contribution
$j_i$). Class $(a)$ events evidently also contribute to the current
across bond $(i-1,i)$.  Of all events that contribute to $J_{i-1,i}$,
class $(a)$ is that subset of events in which site $i$ is fully
occupied. Since the probability of a jump between $i-1$ and $i$ is
independent of the probability of occupation of site $i$, we have
$J_{i,i+1}=J_{i-1,i}~p_i(l_{i}) +j_i$ where $
j_i=\sum_{n_i=1}^{l_i}{\epsilon(n_i|l_i)}~ p_i(n_i) =
\epsilon_0~z~(1-p_i(l_i)) $
and $p_i (n_i) = u_i (n_i) z^{n_i}/Z_i$.  
Noting that $J_{i-1,i}=J_{i,i+1}=J_0$, we find
$J_0=\epsilon_0~ z$.
Since both $J_0$ and $\rho$ are known in terms of $z$, the
macroscopic speed  $c_0=\partial {J_0}/ \partial\rho$ 
of the kinematic wave can be determined \cite{LW}. 

Interestingly, the steady state measure and current can also be found
in $d>1$, for models in which the ratio of hopping rates in different
directions is independent of site and configuration, and the cascade
of adjacent-site overflows in a single move is in the same direction
\cite{unpubl}.

The other model we investigate is the disordered fully asymmetric
simple exclusion process (DASEP) on a 1-$d$ ring. In this model, each
site can hold 0 or 1 particle. A particle hop is attempted to the
nearest-neighbour site on the right, and is completed only if the site
in question is vacant. Attempt rates are associated with bonds and are
disordered, with magnitudes varying from bond to bond, chosen from a
binary distribution.  Unlike in the DDPP, there is no analytical
characterization of the steady state even with a single inhomogeneous
bond. A numerical study has shown that if $\rho$ is close to $1/2$, a
single weak bond acts as a blockage and produces unequal densities
over macroscopic length scales on either side of it, and a shock where
the density profiles meet, far from the blockage \cite{Janowsky}.  We
have studied the disordered case with an extensive number of weak
bonds by Monte Carlo simulation and found that the steady state
depends strongly on the filling.  The current $J_0$ varies smoothly
with $\rho$ provided $|\rho-1/2| >\Delta$ (Regime A), whereas $J_0$
has a single value if $|\rho-1/2| <\Delta$ (Regime B) (Fig. 2a). The
value of $\Delta$ depends on the ratio $r$ of rates of the 
\begin{figure}
\psfig{figure=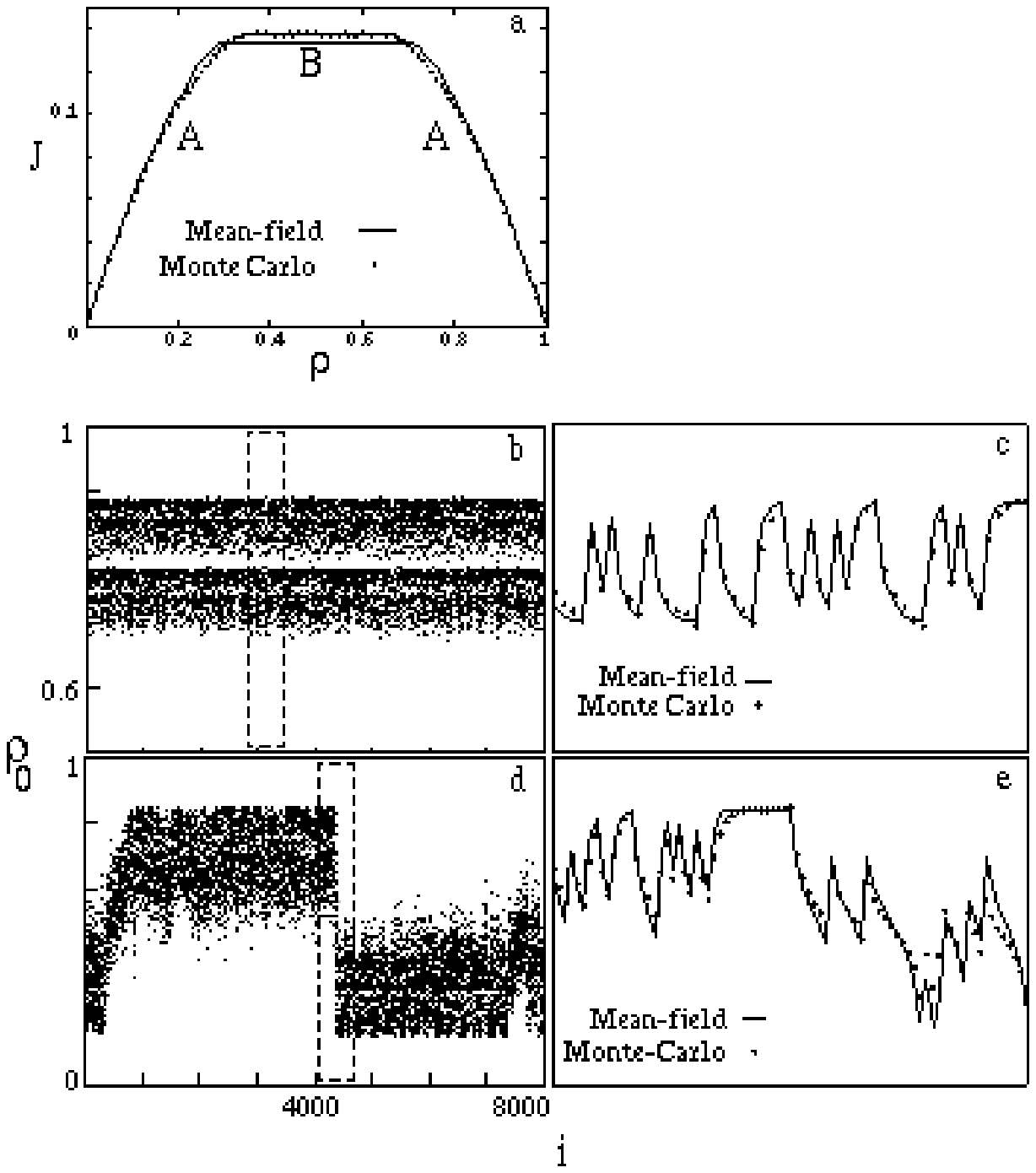,clip=true,width=8cm}
Fig. 2~ DASEP steady states (a) $J$ {\it vs} $\rho$  for $r=f=0.5$ 
(b) density profile for $\rho=0.80$ (c) blowup of  box in (b)
 (d) profile for $ \rho=0.5$ (e) blowup of box in (d).
\label{fig2}
\end{figure}
\noindent weak and strong bonds and on the fraction $f$ of weak bonds, and is $\simeq
0.16$ for $r=f=0.5$.  In Regime A, the density profile consists of a
large number of shocks with a mean inter-shock spacing of a few
lattice spacings (Fig. 2c); the number of shocks scales with the
system size. On length scales large compared to the inter-shock
spacing, the density is roughly uniform.  A semi-quantitative picture
of the steady state can be obtained using a mean-field approximation,
writing the current between sites $i$ and $i+1$ as
$J=\epsilon_{i,i+1}\rho_0(i) (1-\rho_0(i+1))$ where the rate
$\epsilon_{i,i+1}$ is $\epsilon_0$ $(\epsilon_0/2)$ for a strong
(weak) bond. As $J$ is the same in every bond, the densities
$\{\rho_0(i)\}$ satisfy a set of coupled, nonlinear equations, which
can be iterated to convergence.  The result is shown in Fig. 2c. The
mean-field approximation evidently obtains locations of shocks fairly
well, but not shapes of individual shocks.  In Regime B, Monte Carlo
results and mean-field
calculations show that the density is nonuniform on a {\it macroscopic} scale (Fig. 2d),
besides showing shock structure on the scale of a few lattice spacings
(Fig. 2e).  In this regime, the occurrence of long stretches of
successive weak bonds, coupled with the requirement of spatial
uniformity of current, results in phase separation into high and low
density phases, qualitatively as in the single defect case
\cite{Janowsky}. In both regimes, numerical results show $\Gamma(r) \sim
r^\alpha $ with $\alpha\simeq 0.5$

Turning now to the dynamical behaviour of fluctuations in the steady state,
we first consider systems which have a uniform density on macroscopic
scales.  The dynamics of such 1-$d$ systems is dominated by kinematic waves
which transport density fluctuations through the 
system at a mean speed
$c_0$.  Owing to quenched disorder, the local speed of the wave varies from
one location to another (Fig. 3) and the question arises how wave
dissipation, and thus $S(t)$, is affected.  In the pure system, the
evolution equation (\ref {delrho}) has $x$-independent coefficients and the
long-time growth of $S(t)$ is described by $\beta =1/3$ \cite{vanB,KPZ}. In
the disordered system, let us regard the coarse-grained medium as made of
successive disordered patches, each independent of the other, and ask for
the behaviour of a large-scale density fluctuation as it passes through the
succession of patches. If $c_0$ is nonzero, the probability of a density
fluctuation in an infinite system returning to the same disorder patch dies
down rapidly at long times, so it is a good approximation to regard the
effect of disorder as essentially uncorrelated in time.  Further, since the
speed $c(x)$ is a spatially random function, the use of the averaged value
$c_0$ in (\ref{S2}) induces a noise of amplitude $t^{1/2}$ in the location
of the density fluctuation. Since fluctuations in $h$ scale as $x^{1/2}$, 
the effect $(\sim t^{1/2})$ on $S^2$ is subleading. Thus we conjecture that
if $c_0 \neq 0$, the long time behaviour of $S^2(t)$ is $\sim t^{2/3}$, the
same as for the pure system. Our argument differs from that used earlier
for the effect of point disorder on unpinned, moving interfaces
\cite{Parisi}, as our
case corresponds to columnar disorder in the interface
language. The irrelevance of randomness in $c(x)$ is consistent with 
straightforward power counting in (2).

In our numerical determination of $S(t)$ from simulations of the two
types of lattice models, we defined $ h(i,t)$ as
$\sum_{i_0(t)}^{i}\rt(k,t)$ where $i_0 (t)$ is the position of a
specific particle. We averaged results for $S(t)$ over $40$ time
evolutions for a fixed realization of disorder, and then over several
realizations.  For drop-push dynamics, we considered a model with only
two types of wells $A$ and $B$, distributed randomly.  Each well can
hold at most one particle, but the jump rates out of the two types of
wells are different, say $\epsilon_A$ and $\epsilon_B$.  We used
$\epsilon_A/\epsilon_B=0.5$, and a fraction $f=0.5$ of low
rates. Since the placement of the wells is random, the essential
feature of quenched disorder is still present.  Since (\ref {InvM})
still holds, we start with an initial configuration of particles
consistent with this product measure.  The analysis is aided by the fact
that 
\begin{figure}
\psfig{figure=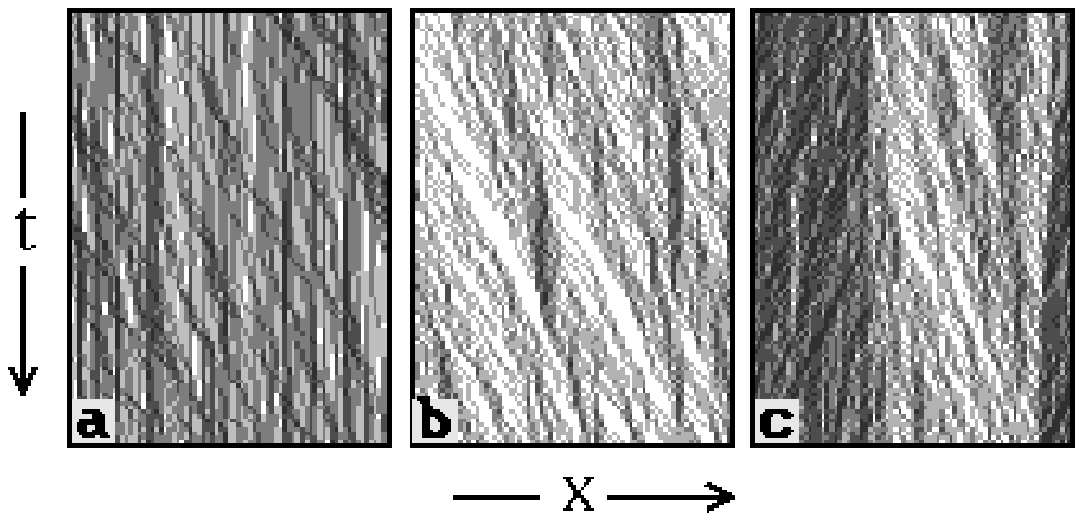,clip=true,width=8cm}
Fig. 3~ Typical time evolutions (a) DDPP (b) DASEP, $\rho=0.25$ 
(c) DASEP, $\rho=0.5$. Darker regions are particle-rich.
The tilted streaks are kinematic waves.
\label{fig3}
\end{figure}
\begin{figure}
\psfig{figure=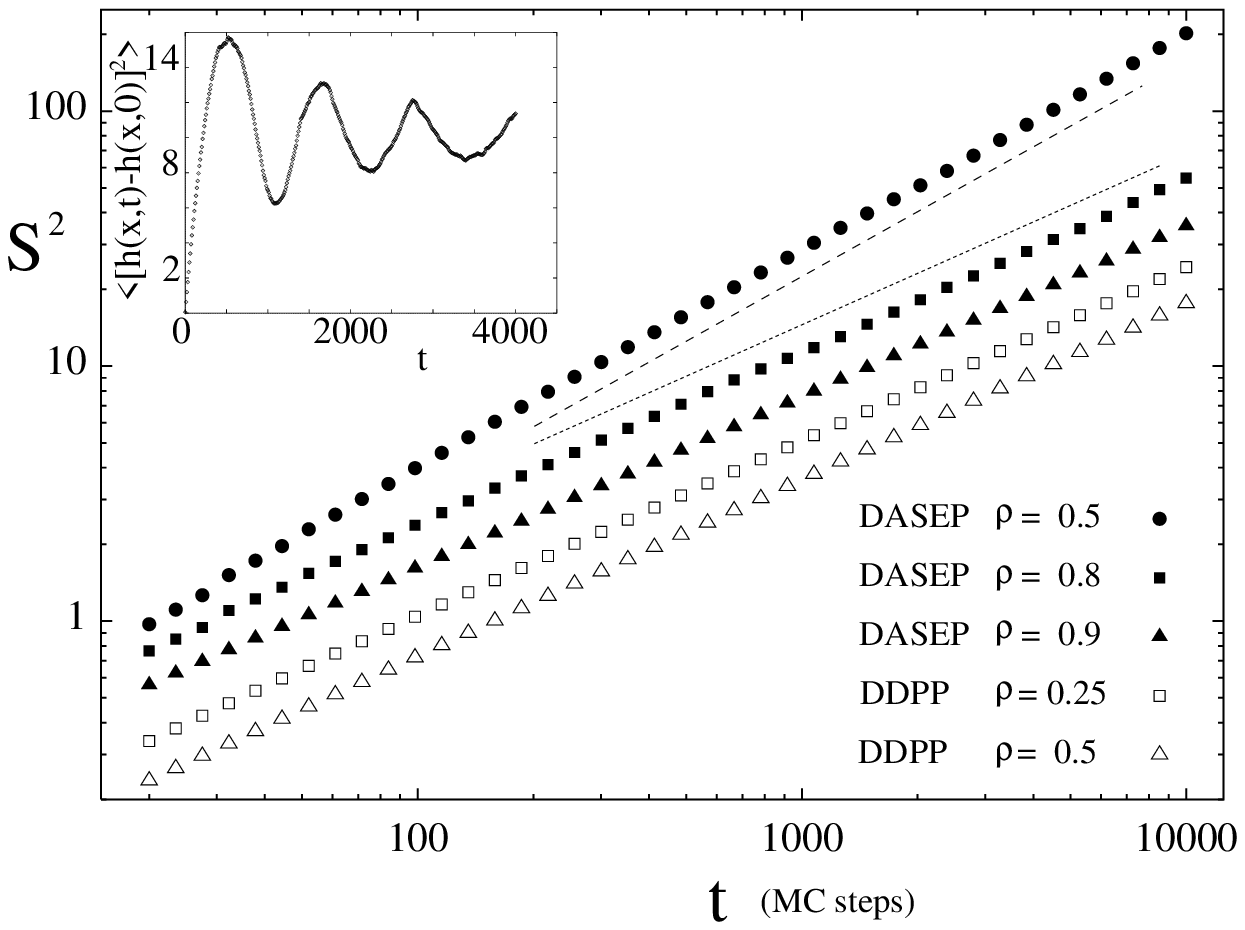,clip=true,width=8cm}
Fig. 4~ $S^2(t)$ for the DDPP and the DASEP for different
  densities. Individual data sets have been shifted for clarity. The
  straight lines have slopes $0.84$ and $0.67$.  Inset: The return
  time of the kinematic waves is given by the period of oscillations
  of the autocorrelation function.
\label{fig4}
\end{figure} 
\noindent $J_0$ and $c_0$ are known explicitly. 
  In the case of the DASEP, we chose $r=f=0.5$ and used a system of size $8000$, allowing
it to relax for $\simeq 60,000$ Monte Carlo steps to achieve steady
state. Further, $J_0$ and $c_0$ were estimated in two stages: First, a
rough estimate of $J_0$ was obtained by direct measurement, while $c_0$ (in
Regime A) was estimated from the return time of the kinematic wave in a
smaller system (Fig. 4 inset). Then a more accurate estimate was obtained
by minimizing $S$ in a large system with respect to $J_0$ and
$c_0$. Details of the minimization procedure will be given elsewhere
\cite{unpubl}.  Fig. 4 shows data for two different densities for the DDPP
and also Regime A of the DASEP, all corresponding to nonzero $c_0$. In all
these cases, the data is consistent with $S^2\sim t^{2/3}$, as for
the pure system, supporting our conjecture. Further, we
performed a direct check of the $\sim t^{1/2}$ growth of
disorder-induced noise in locations of density fluctuations, discussed
in the previous paragraph. Such noise should alter the growth law for
the sliding-tag correlation function \cite{MB} (which monitors
fluctuations of tagged-particle locations) from $\sim t^{1/3}$ to
$\sim t^{1/2}$ -- a change we confirmed by simulation of the DDPP
\cite{unpubl}.

In Regime B of the DASEP, $S(t)$ shows stronger fluctuations than in
Regime A, from one realization of disorder to another. On averaging
over 10 samples, we find $S(t)$ grows as $t^{\beta}$ with
$\beta=0.42\pm0.02$ (Fig. 4).  It is possible that the more rapid
growth of $S$ in this case arises from oppositely moving kinematic
waves in different macroscopic regions (Fig. 3c).

We conclude with a recapitulation of our principal results.  For the DDPP,
the steady state has a product measure form, and the current can be
determined. For the DASEP, the steady state density profile shows many
shocks and is quite well described by a mean-field approximation.  Our
conjecture, that disorder does not affect the dynamical universality class
if the kinematic wave speed $c_0$ is nonzero, is borne out by simulation
studies of the DDPP and the DASEP in Regime A. A vanishing $c_0$ can
indicate phase coexistence with different densities in different
macroscopic regions, as in Regime B of the DASEP; the dynamical behaviour
is different in this case.

We thank R. E. Amritkar, D. Dhar, S. Krishnamurthy and G. I. Menon for
useful comments.

\end{multicols}

\begin{thebibliography}{99}
\bibitem{Fisher}{O. Narayan and D.S. Fisher, Phys. Rev. B {\bf 49}, 9469 (1994)}
\bibitem{LW}{M. J. Lighthill and G. B. Whitham, Proc. R. Soc.
London, Ser. A{\bf 229}, 281 (1955); {\bf 229}, 317 (1955).}
\bibitem{Amaral}{L. A. N. Amaral, A. -L. Barabasi and  H. E.
Stanley, Phys. Rev. Lett. {\bf 73}, 62 (1994).}
\bibitem{Krug}{J. Krug, Phys. Rev. Lett. {\bf 75}, 1795 (1995).}
\bibitem{BR}{M. Barma and R. Ramaswamy in {\it Non-linearity And Breakdown 
in Soft Condensed Matter}, edited by K. K. Bardhan, B. K. Chakrabarti 
and A. Hansen (Springer, Berlin, 1993), p.309.}
\bibitem{SRB} {G. Sch\"utz, R. Ramaswamy and M. Barma, J. Phys. A{\bf
29}, 837 (1996).}
\bibitem{vanB}{H. van Beijeren, R. Kutner and H. Spohn, Phys. Rev.
Lett. {\bf 54}, 2026(1986)}.
\bibitem{KPZ}{M. Kardar, G. Parisi and Y. -C. Zhang, Phys. Rev. Lett.
{\bf 62}, 89  1986.}
\bibitem{MEQ}{ N. G. van Kampen, {\it Stochastic Processes in Physics
and Chemistry}, (North Holland, Amsterdam, 1981).}
\bibitem{traffic}{However, models in which quenched random jump rates are
associated with particles have been considered in M. R. Evans, Europhys. 
Lett. {\bf 36}, 13 (1996) and J. Krug and P. A. Ferrari, J. Phys. A{\bf 29}
 L465 (1996).} 
\bibitem{unpubl}{G. Tripathy and M. Barma, unpublished}
\bibitem{Janowsky}{ S.A. Janowsky and J.L. Lebowitz, Phys. Rev. A{\bf
45}, 618 (1992); G. Sch\"utz, J. Stat. Phys. {\bf 75}, 471 (1993).}
\bibitem{Parisi}{ G. Parisi, Europhys. Lett. {\bf 17}, 673 (1992).}
\bibitem{MB}{S.N. Majumdar and M.Barma, Phys. Rev. B{\bf 44}, 5306
(1991); Physica A{\bf 177}, 366 (1991).}  

\end{thebibliography}
\end{document}